\documentclass[journal=ancac3,manuscript=letter,layout=twocolumn]{achemso}

\usepackage{graphics}      
\usepackage{amsmath}
\usepackage{amssymb}
\usepackage{textcomp}

\makeatletter
\let\l@addto@macro\relax
\makeatother
\usepackage[fontsize=9pt]{scrextend}
\usepackage[font=small,labelfont=bf]{caption}

\newcommand{\nm}{\textrm{nm}}
\newcommand{\cm}{\textrm{cm}}
\newcommand{\um}{\textrm{\textmu m}}

\title{Imaging Metasurfaces based on Graphene-Loaded Slot Antennas}

\author{Jordan A. Goldstein}
\affiliation[Massachusetts Institute of Technology]{Electrical Engineering and Computer Science, Massachusetts Institute of Technology, Cambridge, Massachusetts, United States}
\email{jordango@mit.edu}
\author{Dirk R. Englund}
\affiliation[Massachusetts Institute of Technology]{Research Laboratory of Electronics, Massachusetts Institute of Technology, Cambridge, Massachusetts, United States}


\begin{document}

\begin{abstract}
Spectral imagers, the classic example being the color camera, are ubiquitous in everyday life. However, most such imagers rely on filter arrays that absorb light outside each spectral channel, yielding $\sim\!\!1/N$ efficiency for an $N$-channel imager. This is especially undesirable in thermal infrared (IR) wavelengths, where sensor detectivities are low, as well as in highly compact systems with small entrance pupils. Diffractive optics or interferometers can enable efficient spectral imagers, but such systems are too bulky for certain applications. We propose an efficient and compact thermal infrared spectral imager comprising a metasurface composed of sub-wavelength-spaced, differently-tuned slot antennas coupled to photosensitive elements. Here, we demonstrate this idea using graphene, which features a photoresponse up to thermal IR wavelengths. The combined antenna resonances yield broadband absorption in the graphene exceeding the $1/N$ efficiency limit. We establish a circuit model for the antennas' optical properties and demonstrate consistency with full-wave simulations. We also theoretically demonstrate broadband $\sim\!36\%$ free space-to-graphene coupling efficiency for a six-spectral-channel metasurface. This research paves the way towards compact CMOS-integrable thermal IR spectral imagers.
\end{abstract}

\begin{figure}[th]\centering
\includegraphics[width=\columnwidth]{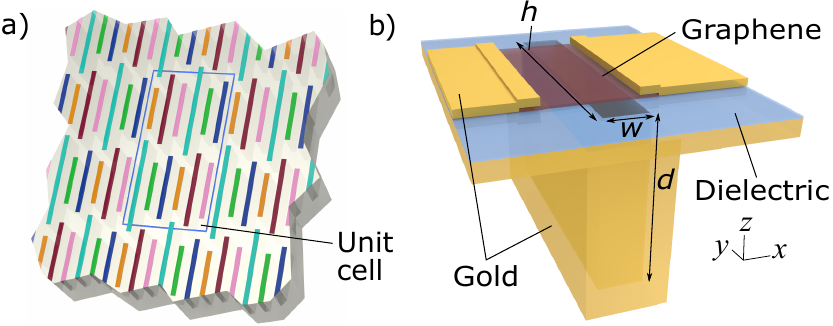}
\caption{a) Illustration of a broadband absorbing slot antenna metasurface consisting of six differently tuned slot antennas tiled with subwavelength periodicity. The graphene patches are color-coded by antenna length, and the diagram is drawn to proportion based on the device dimensions used to produce Figure \ref{antennas6}. b) Depiction of a single graphene-coupled slot antenna-based photodetector based on the photothermoelectric effect.}
\label{figure1}
\end{figure}
\subsubsection{ }
\vspace{-20pt}We take spectral imaging for granted in daily life. Our eyes are spectral imagers, providing information about the composition of what we see. The infrared electromagnetic band covers many chemical absorption resonances and thus also reveals compositional information. In particular, infrared spectral imaging is applied in areas such as gas emission monitoring\cite{emissions,gases,etalonplume,boat,agent,seepage,hawaii}, ecological monitoring\cite{soil,trees,pines}, food quality control\cite{foodreview,foodquality,corn}, waste sorting\cite{plastic}, biological research\cite{dichroic} and oceanography\cite{ocean}.

Spectral imaging aims to measure a ``data cube'' representing light intensity over two spatial dimensions $x$ and $y$ and one spectral dimension $\lambda$ with $N$ channels. Scanning spectral imagers sequentially measure different portions of the data cube over multiple exposures to form the full data cube. A common example is the pushbroom scanner which measures $x \times \lambda$ data cube slices while scanning $y$ and is thus typically associated with satellites and conveyor belts where either the camera or subject is gradually moving in one direction\cite{satellite,emissions,foodquality,plastic}. The spectral axis may also be scanned such as in tunable filter-based imagers \cite{etalonplume,liquidcrystal}, which feature at most $1/N$ light utilization efficiency, or Fourier transform interferometer-based imagers which are bulky and have moving parts\cite{gases}. In contrast, snapshot spectral imagers (SSIs) capture a data cube with a single exposure\cite{snapshot}. This may be achieved using a color filter array similar to that of a color camera, thus limiting efficiency to $1/N$\cite{atwater,nanowires,snapshot}. Another category of SSIs uses dichroic or dispersive optics to break up incoming light by wavelength before arriving on a focal plane array (FPA). There are many variations of this approach\cite{bodkin,coded,ims,gorman,dichroic}, but they all require increasing the etendue of the incoming light beam by a factor of $N$, leading to an unfavorable tradeoff between total FPA area, input acceptance angle and spectral resolution\cite{snapshot}.

Compared to these technologies, the category of imagers based on inherently multispectral pixels is less explored. One such example is the Foveon RGB sensor, which extracts three different electrical signals from different depths in the optically active silicon region, as shorter wavelengths are absorbed closer to the sensor surface\cite{foveon}. Another approach uses nanoantennas, optical resonators of subwavelength dimensions which nevertheless feature absorption cross sections of order $\lambda^2$ if the antenna is conjugate impedance matched with its load; or, equivalently, if the antenna is critically coupled to the vacuum. Here, we propose a thermal IR multispectral imager where $N$ differently sized metallic slot antennas with infrared-sensitive loads targeting $N$ spectral channels are tiled to form a metasurface featuring efficient free space-to-load optical energy transfer. Figure \ref{figure1}a shows such a metasurface for $N=6$. We model graphene as the photosensitive load because its broadband absorption in the mid-IR\cite{grcond} and processing flexibility\cite{grcmoscamera,grcmoschem} make it suitable for this platform. Not only do the antennas sort incident light by spectral channel, but they also enhance the absorption of the graphene load, bridging the gap between the impedance of free space and graphene, which, when undoped, has an optical sheet resistance no lower than $16.1\,\text{k}\Omega$.\cite{grcond}. Figure \ref{figure1}b shows in detail a single such antenna-coupled graphene photodetector. This detector is designed for a strong photothermoelectric response, in which absorbed light heats up the electron gas in the graphene, resulting in an electromotive force due to the Seebeck effect. The graphene channel is assumed to be isolated from the metasurface by a several-nanometer layer of dielectric, thin enough to not impact the optical properties of the system. The asymmetric position of the graphene channel with respect to the slot allows half of the graphene channel to be gated by metal underneath, yielding the asymmetric graphene Fermi level profile needed for a nonzero net photoresponse\cite{gabor,jsong}. Note that while perfect absorption in this wavelength range has been demonstrated in heavily doped graphene accompanied with metal nanostructures\cite{seyoon}, we limit our consideration to undoped graphene as the peak Seebeck coefficient occurs at very low doping levels\cite{gabor}.

For optical absorption, slot antennas offer a few advantages over planar designs such as dipole or bowtie antennas. They have unidirectional radiation patterns, and thus an array of them can perfectly absorb an incident beam, whereas planar antennas require a quarter-wave back-reflector to do so \cite{alaee,eggles}. The wavelength dependence of the back-reflector phase complicates design of broadband absorbing metasurfaces and exacerbates undesirable antenna-antenna coupling. Additionally, since planar antennas must be supported by transparent dielectric, they cannot be embedded in a CMOS process as the inter-layer dielectric strongly absorbs thermal infrared radiation\cite{kischkat,lin}, whereas for slot antennas the dielectric on top and in the perforations may be etched away without sacrificing the mechanical integrity of the antenna.

\subsubsection{Results and Discussion}

\begin{figure}[th]\centering
\includegraphics[width=\columnwidth]{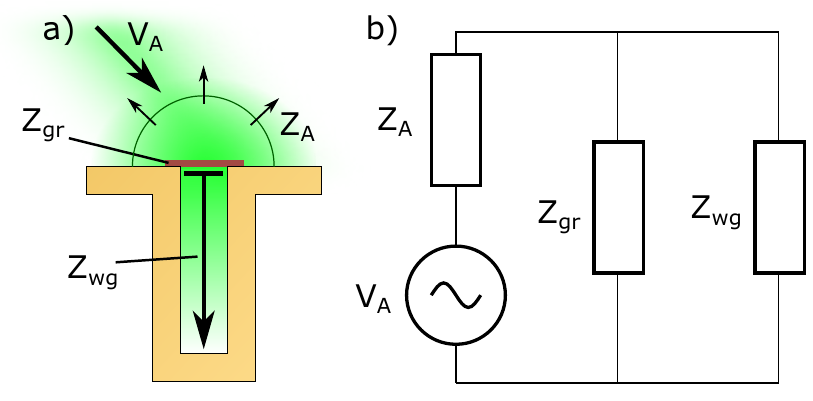}
\caption{a) Graphene-loaded slot antenna with physical features corresponding to the components in circuit b) labelled. b) Circuit schematic of a graphene-coupled slot antenna. $V_A$ represents incoming light, $Z_A$ is the radiation impedance of the slot aperture, $Z_\text{gr}$ is the impedance of the graphene sheet and $Z_\text{wg}$ is that of the slot, effectively a short-circuit waveguide stub.}
\label{figure2}
\end{figure}
We model the slot antenna depicted in Figure \ref{figure2}a as an aperture antenna fed by a rectangular waveguide terminated in a short circuit a distance $d$ behind the aperture. We represent the graphene sheet as a shunt impedance connected in parallel with the waveguide stub. Figure \ref{figure2}b illustrates this circuit with a Th\'{e}venin equivalent radiation impedance $Z_A$ and source $V_A$ representing the aperture antenna\cite{stutzman}. The rectangular waveguide stub presents an impedance 
\begin{equation}
Z_\text{wg} = Z_0\,\frac{e^{jn_\text{eff}k_0d} + r e^{-jn_\text{eff}k_0d}}{e^{jn_\text{eff}k_0d} - r e^{-jn_\text{eff}k_0d}},
\end{equation}
where $Z_0$ and $n_\text{eff}$ are the characteristic impedance and effective index of the $\text{TE}_{10}$ mode of the slot and $k_0$ is the vacuum wavenumber. $r$ is the Fresnel reflection coefficient between vacuum and metal for an \emph{s}-polarized plane wave at an incident angle of $\arccos(n_\text{eff})$, which describes the $\text{TE}_{10}$ mode.

The graphene presents a mostly resistive impedance of
\begin{equation}
Z_\text{gr}=\frac{\pi^2w}{8h\sigma_\text{gr}},
\label{zgr}
\end{equation}
using power/current impedance normalization\cite{schelkunoff}. $w$ and $h$ are defined in Figure \ref{figure1}b and $\sigma_\text{gr}$ is the sheet conductance of the graphene, modeled here as intrinsic. We calculate $Z_A$ using finite element simulations for various $h$ and $w$; we provide more details on these calculations in the Methods section. See Supplementary Figure 1 for an example of the frequency dependence of the impedances in the circuit.

Define $\eta_\text{gr}$ as the fraction of available power from Th\'evenin source $V_A$, $Z_A$ that is dissipated in $Z_\text{gr}$. Solving the above circuit, we arrive at
\begin{equation}
\eta_\text{gr}=4\left|\frac{Z_\text{gr}\parallel Z_\text{wg}}{Z_\text{gr}}\right|^2\frac{\operatorname{Re}(Z_A) \operatorname{Re}(Z_\text{gr})}{\left|Z_A+(Z_\text{gr}\parallel Z_\text{wg})\right|^2},
\label{etaeqn}
\end{equation}
where $\parallel$ represents reciprocal addition. Define $A_\text{gr}=\frac{P_\text{gr}}{I_\text{inc}}$ as the partial absorption cross section of light of intensity $I_\text{inc}$ coupled into the graphene and $P_\text{gr}$ as the power absorbed in the graphene. For a lossless, conjugate matched antenna, antenna theory predicts $A_\text{gr,max}=\frac{D\left(\theta,\phi\right)}{4\pi}\lambda^2$ where $D\left(\theta,\phi\right)$ is the antenna's directivity at the given incident angle and polarization\cite{stutzman}, which we calculate using finite element simulations. $\theta$ here is the polar angle and $\phi$ is the azimuthal angle. We omit the polarization angle in our notation, implicitly setting it to maximize $D$. Additional ohmic losses and impedance mismatch then reduce the actual absorption cross-section into graphene by a factor $\eta_\text{gr}$, i.e.
\begin{equation}
A_\text{gr}=\eta_\text{gr}\frac{D\left(\theta,\phi\right)}{4\pi}\lambda^2. 
\label{agr}
\end{equation}
We obtain $A_\text{gr}$ from FDTD simulations of plane waves incident on the graphene-loaded antennas, which we then use to calculate $\eta_\text{gr}$ via Equation \ref{agr}.

\begin{figure}[ht]\centering
\includegraphics[width=\columnwidth]{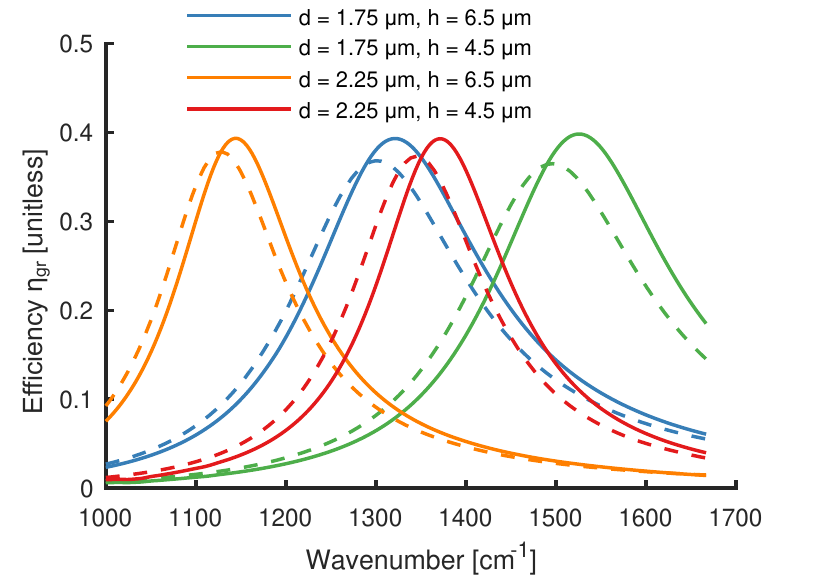}
\caption{Comparison of simulated and modeled $\eta_\text{gr}$. Modeled antennas are $400\,\text{nm}$ wide. Dashed lines represent the fullwave FDTD absorption results, while the solid lines represent the impedance model results.}
\label{efficiencies}
\end{figure}

Fig. \ref{efficiencies} compares $\eta_\text{gr}$ between the model described by Equation \ref{etaeqn} and FDTD results for antennas of various dimensions. The data show that the model is accurate to within 10\% of the $\eta_\text{gr}$ peak amplitude and 2\% of the resonance wavenumber. We attribute these discrepancies to aspects not captured by the quasi-analytical model, such as our assumption of a perfectly conducting outer antenna face and finite meshing. Despite these shortcomings, the present model allows us to predict slot antenna absorption properties to tolerances comparable to the uncertainty due to variations in metal quality.

\begin{figure}[ht]\centering
\includegraphics[width=\columnwidth]{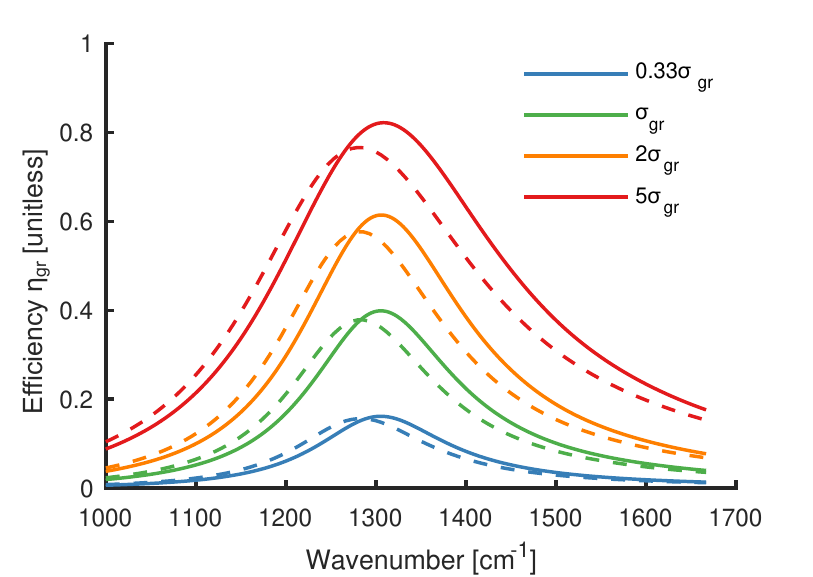}
\caption{$\eta_\text{gr}$ versus wavenumber for various values of the load sheet conductance, where $\sigma_\text{gr}$ is the optical conductivity of intrinsic graphene. Dashed lines represent the fullwave FDTD absorption results, while the solid lines represent the model results. The antenna featured here has dimensions $d = 2.0\,\um$, $h = 5.5\,\um$, and $w = 0.4\,\um$. }
\label{grcondvar}
\end{figure}

To further validate our model, we artificially scale the sheet conductance of the graphene load by factors ranging from $0.33$ to $5$ and compare the resulting $\eta_\text{gr}$ between the model and FDTD for a single such antenna. The results, shown in Figure \ref{grcondvar}, show that our model accurately predicts the sublinear scaling of $\eta_\text{gr}$ with respect to the load conductivity, with the $\eta_\text{gr}$ peak amplitude reaching $0.8$ for a load conductivity of $5\sigma_\text{gr}$.

Having modeled the individual components, we now discuss broadband absorbing metasurfaces incorporating differently tuned antennas tiled in a periodic array. We use a three-step process to design such metasurfaces. We first constrain $d$ and $w$ to be the same for all antennas in the metasurface, and choose their values to yield high peak $\eta_\text{gr}$ for antennas resonant in the targeted wavelength range. Secondly, we choose the values of $h$ for the antennas, following the heuristic that at the wavelength where one antenna's $\eta_\text{gr}$ falls to half its peak amplitude, the next antenna's $\eta_\text{gr}$ should have risen to half its peak amplitude. Finally, we choose the arrangement and pitch of the antennas to be as closely packed as possible while satisfying qualitative fabricability constraints. We also avoid juxtaposing antennas of adjacent wavelength channels to minimize antenna-antenna crosstalk.

The antenna pitch, more accurately described by the Bravais lattice vectors of the array, is a critical parameter in determining the potential absorption efficiency of the array. Light incident from a given direction can only be scattered by the two-dimensional diffraction orders of the lattice. The array can only perfectly absorb an incoming light beam if no nonzero diffraction orders fall within the light cone, barring the event that all higher diffraction orders overlap with nodes in the individual unit cell radiation pattern. By ``light cone'', we refer here to the region in the Fourier transform space of the $xy$--plane for which radiation can occur, namely $k_x^2 + k_y^2 < k_0^2$.  For a square lattice, if the lattice pitch $a<\lambda_\text{min}/2$, no higher diffraction orders are within the light cone for any incident angle. In practice, the limited numerical aperture of imaging systems relaxes this constraint. 

\begin{figure}[ht]\centering
\includegraphics[width=\columnwidth]{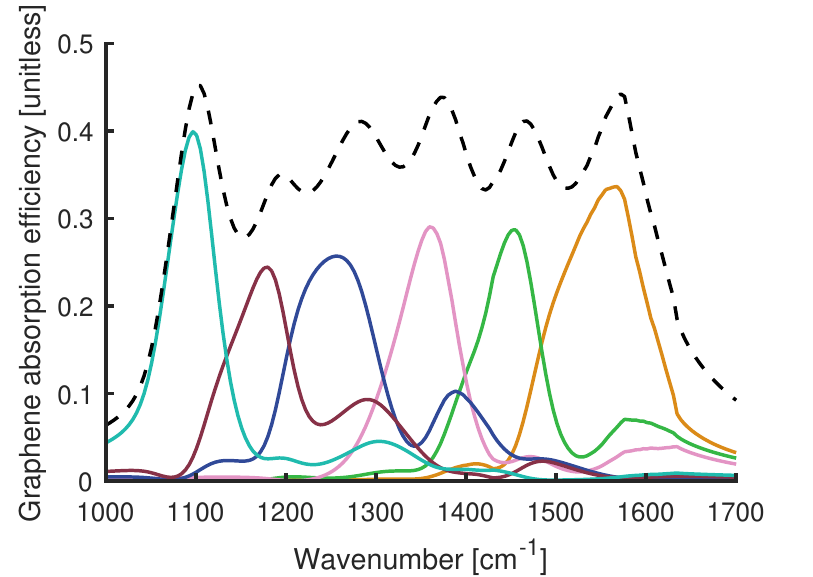}
\caption{Graphene absorption efficiency of the six-antenna metasurface. The colored curves represent the contributions of each antenna to the overall graphene absorption of the metasurface, which is represented by the black dashed curve. The curve colors here match the antenna colors in Figure \ref{figure1}a. Shorter antennas yield higher resonance wavenumbers.}
\label{antennas6}
\end{figure}

Targeting the $6-10\,\um$ wavelength range, we follow the above methodology and arrive at an array of six antennas with $d=2.25\,\um$, $w=400\,\nm$, and $h={3.41, 3.81, 4.41, 5.04, 5.80, 7.36}\,\um$, uniformly spaced and tiled as shown in Figure \ref{figure1}a with a $7\,\um$ by $12.667\,\um$ unit cell. Figure \ref{antennas6} shows the absorption efficiency of normally incident light into the graphene load of each antenna as well as their sum, simulated with FDTD. With an average efficiency of $36\%$ across the $1050\,\cm^{-1}$ to $1600\,\cm^{-1}$ band, this structure improves upon the $1/N$ limit of filter array-based imagers by roughly a factor of two. Note that the unit cell highlighted in Figure \ref{figure1}a is not the primitive unit cell of the lattice, although it was used as the FDTD simulation region due to software constraints.

\begin{figure}[ht]\centering
\includegraphics[width=\columnwidth]{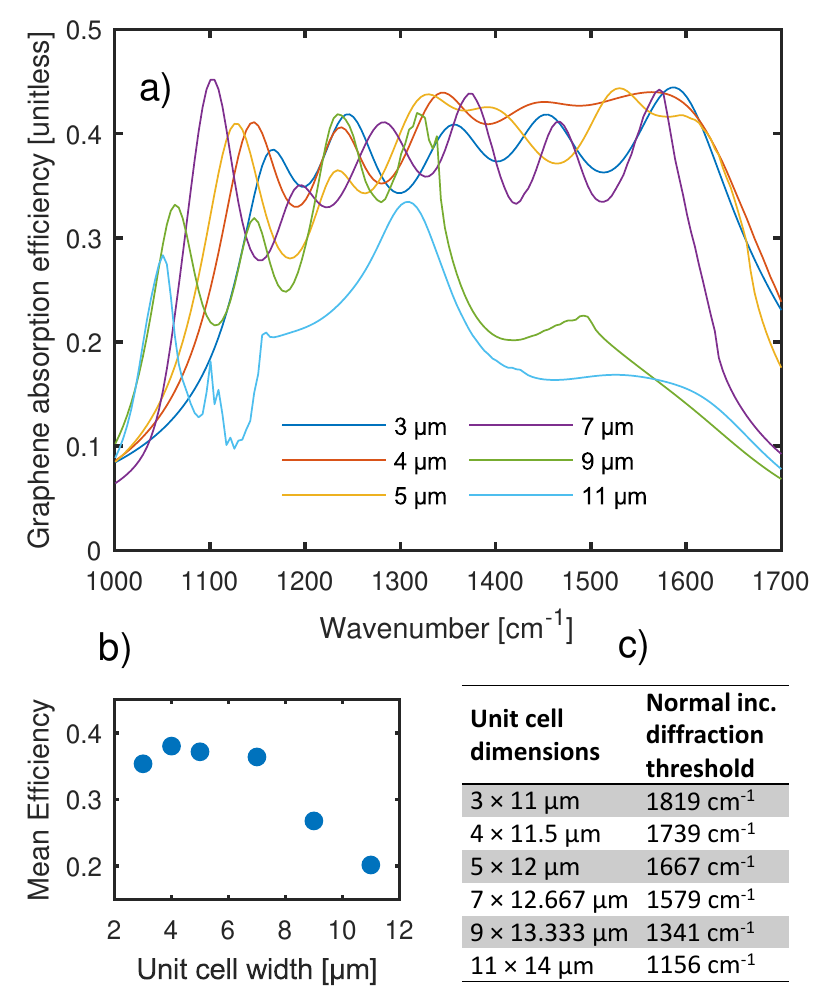}
\caption{a) Total graphene absorption efficiency versus wavenumber for metasurfaces with varied unit cell pitch. The legend indicates the width of the unit cell as illustrated in Figure \ref{figure1}a. b) Mean absorption efficiency for each of the curves in a), averaged between $1050\,\cm^{-1}$ and $1600\,\cm^{-1}$. c) Unit cell dimensions used to obtain the results in a) and b) and the corresponding maximum wavenumbers for which normally incident light experiences no higher order diffraction. The unit cell x and y pitch were varied together to attain a roughly uniform antenna-antenna proximity.}
\label{pitchsweep}
\end{figure}

To further understand the physics of these metasurfaces, we vary the pitch of the unit cell while keeping all other parameters constant. The resulting total absorption efficiency spectra for six different cases are shown in Figure \ref{pitchsweep}a, and their mean efficiencies averaged between $1050\,\cm^{-1}$ and $1600\,\cm^{-1}$ are plotted in Figure \ref{pitchsweep}b. We show the individual absorption contributions from each antenna for each metasurface pitch in Supplementary Figure 2, and we list the unit cell widths and heights in Figure \ref{pitchsweep}c. The data show that the absorption efficiencies are roughly constant, and comparable with the peak efficiencies obtained in Figure \ref{efficiencies}, up to the $7\,\um$--wide unit cell. For larger unit cells, the mean efficiency decreases with increasing unit cell size. This can be understood by analyzing the diffraction characteristics of the various metasurfaces. Sparser metasurfaces have tighter reciprocal lattices and thus more diffraction orders are available within the light cone. For normally incident illumination as we are using here, the minimum wavelength for which no higher-order diffraction occurs (i.e., completely specular reflection) is given by $\lambda_c=\max\!\left(\left(a_1^{-2}+a_2^{-2}\right)^{-1/2},\frac{a_1}{2},\frac{a_2}{2}\right)$, where $a_1$ and $a_2$ are the unit cell dimensions. The corresponding wavenumbers for the various unit cells used here are listed in Figure \ref{pitchsweep}c. For the $9\,\um$ and $11\,\um$ width unit cells, the threshold wavenumber falls in the middle of the range of interest, resulting in reduced efficiency as diffracted light cannot participate in destructive interference with specularly scattered light. This is especially apparent for the $9\,\um$ unit cell, which exhibits a clear spectral transition between high and low efficiency at the diffraction threshold. Note that choosing an excessively tight antenna spacing is also detrimental, not only due to fabrication difficulty, but also because graphene detectors feature minimum Johnson noise-dominated noise-equivalent power for channel lengths comparable to the hot carrier cooling length, which can range from $100\,\nm$ to over $1\,\um$ depending on the substrate and graphene quality \cite{cosmi,jsong,gabor}. Exceedingly tight antenna spacings may not provide room for such long graphene channels.

For a spectrally sensitive metasurface to be practical, not only must it maintain a high absorption efficiency for a reasonable range of incoming light directions, but also the absorption spectra of the individual antennas must not shift or distort too strongly as the incoming light angle varies. The directional dependence of our metasurface arises from two factors: The directivity profile $D\left(\theta,\phi\right)$ of the individual antennas, and array effects resulting from interference and antenna-antenna coupling. Although full angle-dependent simulation results for our gold metasurfaces are outside the scope of this paper due to the extreme computational overhead of off-angle periodic structure simulations\cite{bfast}, we can still provide insight by elaborating on the aformentioned factors, and we also perform off-angle simulations of a simplified metasurface constructed of Perfect Electrical Conductor (PEC) which permits a much larger mesh size. Supplementary Figures 3a, b and c display the directivity profile of a $6.5\,\um\times 400\,\nm$ antenna on resonance. This profile is similar to those of the other antenna lengths used in the metasurface. Intuitively, the directivity decreases as the incident angle approaches the long axis of the antenna, and we thus expect a similar trend in the directional dependence of the array. In Supplementary Figure 3d, we plot for three wavenumbers the sets of incident light directions, represented as components $k_x$, $k_y$ of the incident wavevector $\mathbf{k}$, for which only specular reflection from the metasurface occurs. For $\lambda^{-1}=1050\,\text{cm}^{-1}$, all light incident within $45^\circ$ of normal is only specularly reflected. This range decreases with increasing wavenumber until normally incident light is pinched off at $1579\,\text{cm}^{-1}$. As in Figure \ref{pitchsweep}, we expect efficiency to suffer when the specular reflection-only condition is not met. Supplementary Figure 4 shows the results of off-angle simulations of the simplified PEC metasurface. The data show that for light incident off-angle but perpendicular to the antennas' long axes, antenna resonances falling outside the specular reflection-only region are subject to decreased absorption efficiency as well as blue-shifting. For light incident off-angle and perpendicular to the antennas' short axes, we obtain similar results, except that the peaks red-shift instead of blue-shift, and we observe an overall reduction of the absorption at steep incident angles due to the reduced directivity.

We also investigate metasurfaces comprising numbers of spectral channels besides $N=6$. Supplementary Figure 5 shows the geometric details and simulation results for metasurfaces with $N=3$, 4, 5, and 8 as well as the default value of 6. We achieve good results with uniformly high absorption efficiency for $N=5$ and $6$. For lower values of $N$, the wide frequency spacing between the individual resonances yields deep troughs in the overall absorption efficiency curve, whereas for higher values of $N$, excessive overlap between the antenna resonances causes the overall metasurface efficiency to suffer.

\begin{figure}[ht]\centering
\includegraphics[width=\columnwidth]{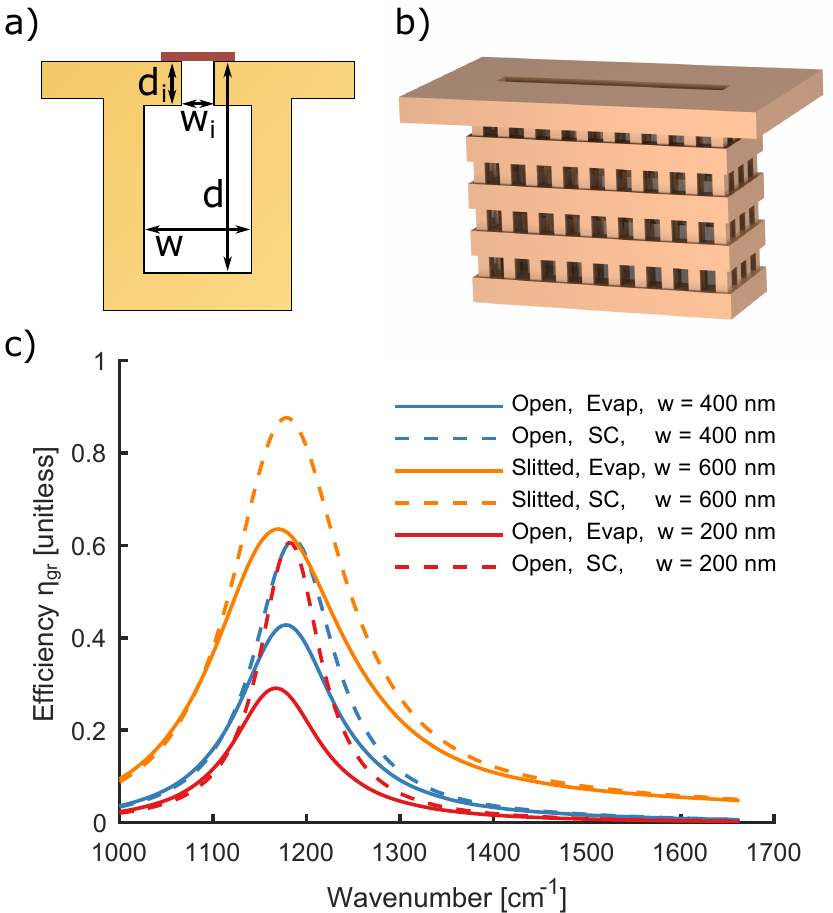}
\caption{a) Cross-section of slot antenna with narrowed input slit, as well as dimension definitions. b) Illustration of how such an antenna could be implemented in the wiring layers of a CMOS chip. c) $\eta_\text{gr}$ versus wavenumber for various antenna geometries and gold optical models. ``Open'' refers to the the normal slot antenna design in Figure \ref{figure1}b), while ``slitted'' refers to the slitted design in a). ``Evap'' refers to evaporated gold as used throughout the paper\cite{palik}, whereas ``SC'' refers to single crystal gold \cite{crc}. For the open antennas plotted here, $d=2.75\,\um$. For the slitted inlet antennas, $d=1.75\,\um$, $w_\text{i}=0.15\,\um$, and $d_\text{i}=0.2\,\um$. These simulations use solid, not perforated sidewalls.}  
\label{improvedgeometries}
\end{figure}

With realistic metal, the efficiency of these devices is ultimately limited by ohmic losses. However, more advanced antenna designs can be used to improve $\eta_\text{gr}$. As it turns out, not only can the copper layers in the back end of a CMOS process be designed to incorporate slot antennas, but they also provide a convenient medium to realize those antenna designs that would be exceedingly difficult to fabricate in an academic setting\cite{gupta}. Figure \ref{improvedgeometries}a introduces such a design in which the slot inlet is narrower than the internal cavity width. This design concentrates the electric field around the graphene, which reduces $Z_\text{gr}$ without increasing the $\text{TE}_{10}$ mode loss. Figure \ref{improvedgeometries}b shows a CMOS adaptation of this design, in which the walls of the cavity are perforated to comply with via design rules. If the perforations are sub-cutoff for the resonant wavelength and sufficiently deep, they do not leak light. Additionally, it would be necessary to etch away the inter-layer dielectric to prevent light absorption.

Besides different antenna designs, material quality also affects efficiency considerably. We explore both of these variations in Figure \ref{improvedgeometries}c, which plots $\eta_\text{gr}$ versus wavenumber for intrinsic graphene as simulated by FDTD for narrowed inlet (``slitted'') antennas and normal slot antennas, as well as with a single-crystal gold model \cite{crc} in addition to our default evaporated gold model. The data show that adopting a slitted antenna design increases the peak $\eta_\text{gr}$ from $0.4$ to $0.6$, and then to $0.9$ for single-crystal gold. We can thus hope to achieve $\eta_\text{gr}\approx0.6$ for CMOS-integrated antennas, as copper has been shown to exhibit slightly superior plasmonic properties to gold given suitable deposition conditions\cite{plasmonicfilms}. While one can achieve high efficiencies with clever antenna designs and more opaque loads than monolayer graphene, the Q is ultimately bounded by that of a sealed metal cavity which we simulate to be about 40 for the present gold model and antenna shapes. Silver has less mid-IR optical loss than copper or gold, but unlike copper it is not considered CMOS-compatible and thus may be difficult to integrate. Polar materials supporting optical phonons in the mid-IR are reported to have high Q plasmonic resonances and are thus worth investigating if higher Q is necessary, although plasmonic behavior only occurs in narrow wavelength bands, limiting applicability\cite{midirplasmonics}.

We can apply data describing the resistivity of gated graphene to our model to estimate the room temperature detectivity of the spectral imager described by Figures \ref{figure1} and \ref{antennas6}.  Using the methods described in Song et al.\cite{jsong} and the electrical properties of polycrystalline, non-annealed graphene achieved by de Fazio et al.\cite{defazio}, we arrive at the values in Table \ref{sensitivity} with and without accounting for graphene contact resistance for normally incident $x$-polarized light. Using the more advanced antenna designs shown in Figure \ref{improvedgeometries}, the detectivities would reach the $10^8\,\text{Jones}$ range. We discuss these calculations in the Methods section. For comparison, more conventional bolometer-based thermal IR FPAs operating at room temperature have been reported to achieve detectivities in the $\text{1--2}\times10^9\,\text{Jones}$ range, although such devices do not feature spectral resolution and are limited to millisecond-range response times\cite{foote,nanovox}.

\begin{table}[t]
\centering
\begin{tabular}{c | c c c}
 $R_\text{c}\;[\Omega\,\um]$& $\mathcal{R}\;[\text{A}/\text{W}]$ & NEP [$\text{pW}\,\text{Hz}^{-\frac{1}{2}}$] & $D^*\,\text{[Jones]}$
 \\\hline\\[-10pt]
0 & $0.83$ & $9.6$ & $6.9\times10^7$ \\ 
1000 & $0.42$ & $12.$ & $5.5\times10^7$ \\
\end{tabular}
\caption{Estimated sensitivity figures for graphene imaging array.}
\label{sensitivity}
\end{table}

Finally, we would like to emphasize the general applicability of the antenna metasurface concept to other wavelength ranges, photosensitive elements and antenna designs. Indeed, Tamang et al. have proposed a similar concept applied to RGB imaging where silicon nanorods act as both the antenna and sensitive element \cite{tamang}. We propose that besides graphene and other 2D materials, III-V or HgCdTe semiconductor photosensitive elements could also be incorporated by a transfer printing heterointegration process\cite{transferprinting}. The slot antenna-based metasurface imager approach could also scale to terahertz, where Ohmic losses are much less than in the mid-IR\cite{ordal} and the antennas could be fabricated directly in a circuit board-like platform.

\subsubsection{Conclusion}

In conclusion, we introduced a six-spectral-channel graphene-coupled slot antenna metasurface with $36\%$ efficiency functioning as a spectral imager in the thermal IR, as well as a model for estimating the optical properties of the individual antennas therein. This device is appropriate for integration in the wiring layers of a CMOS process with suitable post-processing to remove inter-layer dielectric within the cavity and transfer graphene. We have shown that more sophisticated antenna designs can improve the efficiency of optical energy transfer to the load to above $\eta_\text{gr}=0.6$. Further research on this concept may focus on experimental demonstration of the absorption enhancement functionality, or on optimizing the device design to meet certain engineering goals. 

\subsubsection{Methods}

\emph{Simulation details}\newline\vspace{-6pt}\newline
\noindent Unless otherwise specified, we use the evaporated gold optical model described in Palik et al. throughout the paper \cite{palik}. We model graphene as an infinitely thin conductive sheet using the optical conductivity model described in Hanson \cite{hanson}. As input parameters to the model, we use a temperature of $300\,\text{K}$, intrinsic graphene (zero Fermi level), and a scattering rate $\Gamma = 0.514\,\text{meV}$. 

We use the Lumerical FDTD package for our FDTD simulations. For simulations of individual antennas, we use $x$-, $y$- and $z$-meshes of $8\,\nm$ in the vicinities of the slot aperture and slot bottom, as well as a $z$-mesh of $40\,\nm$ within the slot. As such, the finest meshes coincide with metallic surfaces and corners, allowing us to capture the nonzero skin depth of the metal, whereas the more gradual $z$-dependence of the fields inside the slot permits a coarser mesh. We find that a minimum mesh size of $8\,\nm$ yields converged results for these simulations. We use PML boundary conditions on all sides except for the $-z$ side, where we use a metallic boundary condition as the light does not penetrate beyond the slots anyway. We also apply symmetry conditions across the $xz$ and $yz$ planes. For the metasurface simulations, we use the same meshing scheme, but with a fine mesh of $15\,\nm$ and a $z$ mesh of $100\,\nm$ within the slot due to computational resource availability limits. To illustrate the error associated with this choice of mesh, we plot mesh-dependent absorption efficiency curves for a $9\,\um$ by $13.333\,\um$ unit cell, $N=6$ metasurface in Supplementary Figure 6, with the mean efficiency averaged between $1050\,\cm^{-1}$ and $1600\,\cm^{-1}$ plotted in the inset. The results validate our qualitative conclusions and put an approximately $3\%$ relative error bound on the spectrally averaged efficiencies of the metasurfaces, although the coarse mesh does somewhat distort the actual spectra. For the metasurface simulations, we change the $x$- and $y$- boundary conditions to Bloch boundary conditions to reflect the periodic nature of the metasurface, and we apply symmetry across only the $xz$ plane.\newline\vspace{-6pt}

\noindent\emph{Impedance model details}\newline\vspace{-6pt}\newline
\noindent We calculate $Z_0$ and $n_\text{eff}$ for our impedance model using the Lumerical MODE waveguide mode solver with an $8\,\nm$ mesh. We use the graphene model from Falkovsky \cite{grcond}, which gives almost identical results to the Hanson model used by Lumerical for the parameters we use. We use Ansys HFSS finite element software to calculate $Z_A$. These simulations excite the aperture from within by its $\text{TE}_{10}$ mode yielding the $S_{11}$ scattering coefficient of the internally reflected light, from which the software calculates $Z_A$. In the finite element simulations, we model the aperture and slot as perfect electrical conductors, as we expect the real part of the antenna impedance to be dominated by radiative loss (rather than ohmic loss) and the imaginary part by energy storage in the near field of the aperture (rather than plasmonically within the metal). From these same simulations we also extract the antenna directivity $D\left(\theta,\phi\right)$.\newline\vspace{-6pt}

\noindent\emph{Detectivity estimation}\newline\vspace{-6pt}\newline
\noindent We base our estimation of the detectivity $D^*$ on the formulation put forth in Song et al.\cite{jsong}. To calculate the electronic temperature profile of graphene suspended over the slot, we solve the 2-dimensional partial differential equation:
\begin{equation}
-\nabla\cdot\left(\kappa\Delta T_\text{el}\right) + \gamma C_\text{el}\Delta T_\text{el}=\alpha\epsilon_0\dot{N}-\mathbf{j}\cdot\nabla\Pi
\label{diffeq}
\end{equation}
Here $\kappa$ represents the electronic planar thermal conductivity of the graphene; $\Delta T_\text{el}$ is the difference between the thermally excited electronic temperature $T_\text{el}$ and the lattice temperature $T_0=300\,\text{K}$, $\gamma$ represents the electron-phonon thermal decay rate, and $C_\text{el}$ represents the electronic heat capacity of graphene. $\alpha$ is the efficiency with which optical energy absorbed by the graphene is deposited into the electronic system on a sub-picosecond timescale, taken to be unity since the incident photon energy is below graphene's optical phonon energy. $\epsilon_0\dot{N}$ represents the intensity profile of absorbed light. $\mathbf{j}$ is the electrical current density, and $\Pi$ refers to the Peltier coefficient. We choose an antenna with $w = 400\,\nm$ and $h=5.5\,\um$ in these calculations, extracting the spatial dependence of $\epsilon_0\dot{N}$ from Ansys HFSS finite element simulations. For the graphene's conductivity $\sigma$ as a function of Fermi level, we use the data measured by de Fazio et al. for unannealed, polycrystalline graphene \cite{defazio}; this is then used to calculate $\kappa$ via the Wiedemann-Franz law and $\Pi$ as well as the Seebeck coefficient $S$ via the Mott formula. The value of $\gamma C_\text{el}$ is estimated by assuming a electronic thermal cooling length of $\sqrt{\kappa/\gamma C_\text{el}}=1\,\um$, an empirical value\cite{cosmi}. As shown in Figure \ref{figure1}b, the graphene is assumed to be terminated at the slot edge on one side, and is taken to extend $400 \nm$ past the slot edge on the other side, where its Fermi level is gated through the metal to the n-type Seebeck coefficient peak. The graphene Fermi level in the suspended region is simply taken to be the zero-gate-voltage Fermi level from de Fazio et al. as it cannot be controlled. For simplicity we assume a sharp jump in the values of $\sigma$, $\kappa$, $\Pi$ and $S$ between the n- and p-doped sides of the device, neglecting fringing fields from the gate. The graphene channel is assumed to be short-circuited with graphene-metal contact resistances $R_\text{c}$ of either $0$ or $1000\,\Omega\,\um$ per contact, the latter being consistent with 1-dimensional contacts to graphene near the Dirac point\cite{contacts}. The average $\Delta T_\text{el}$ at the graphene p-n junction and the Seebeck coefficients on either side determine the thermal electromotive force via $\mathcal{E}=-S\nabla T$\cite{grosso}, which in turn determines $\mathbf{j}$ via the total device resistance. Thus, Equation \ref{diffeq} including the Peltier term may be solved directly, as $\mathbf{j}$ is a linear functional of $\Delta T_\text{el}$. Solving for $\Delta T_\text{el}$ over two spatial dimensions, we find linear thermal decay profiles along the $+x$ and $-x$ directions away from the $\Delta T_\text{el}$ peak which indicates that the device is in the short-channel regime where carrier cooling is dominated by the $\Delta T_\text{el}=0$ boundary conditions, and $\sqrt{\kappa/\gamma C_\text{el}}$ is large enough for the electron-phonon interaction term to be inconsequential. 

Having obtained the short-circuit responsivity $\mathcal{R}$ under zero bias as such, we calculate the noise-equivalent power of the device assuming Johnson noise at $300\,\text{K}$ as the dominant noise source, a reasonable assumption for an unbiased device\cite{graphenejohnsonnoise}. The detectivity for the array is calculated incorporating the antenna pitch, noting that there are two antennas per unit cell. To account for the decreased optical absorption efficiency of a metasurface loaded with graphene doped to the Seebeck coefficient peaks of roughly $\pm0.05\,\text{eV}$, we redo the simulation used to generate Figure \ref{antennas6} with the graphene doped as such. We plot the resulting absorption spectra in Supplementary Figure 7, which shows a mean absorption efficiency of $33\%$ averaged between $1050\,\cm^{-1}$ and $1600\,\cm^{-1}$. We finally calculate the external values of responsitivity, NEP, and detectivity by scaling the internal values by this efficiency factor.

\subsubsection{Acknowledgement}

The authors would like to thank Chris Panuski and Dr. Laura Kim of MIT as well as Sebastian Castilla of Institut de Ci\`encies Fot\`oniques (ICFO) for helpful feedback and insight in the course of preparing this article. This research was funded in part by a grant from the Institute of Soldier Nanotechnologies of MIT. This material is based upon work supported by the National Science Foundation Graduate Research Fellowship Program under Grant No. 1122374. Any opinions, findings, and conclusions or recommendations expressed in this material are those of the author(s) and do not necessarily reflect the views of the National Science Foundation.

\subsection{Supporting Information}

Additional plots to supplement the main text, including:
\begin{enumerate}
\item Example wavenumber dependence of the impedances used in Eqn. \ref{etaeqn};
\item Graphene absorption efficiency spectra and individual antenna components for metasurfaces of various unit cell pitches;
\item Antenna radiation pattern and diagram of specular reflection-only condition for off-angle excitation;
\item Off-angle graphene absorption spectra for simplified PEC metasurface model;
\item Graphene absorption spectra for metasurfaces with varied numbers $N$ of different antenna sizes;
\item Graphene absorption spectra for a metasurface simulated with different FDTD mesh sizes;
\item Graphene absorption spectra for a metasurface loaded with graphene doped to $0.05\,\text{eV}$. 
\end{enumerate}

\bibliography{antennatheoryacsphot}

\end{document}


\maketitle
\newpage

\subsubsection*{Example of impedances in antenna properties calculation}

\begin{figure}[ht]\centering
\subfloat{{\includegraphics[width=0.65\textwidth]{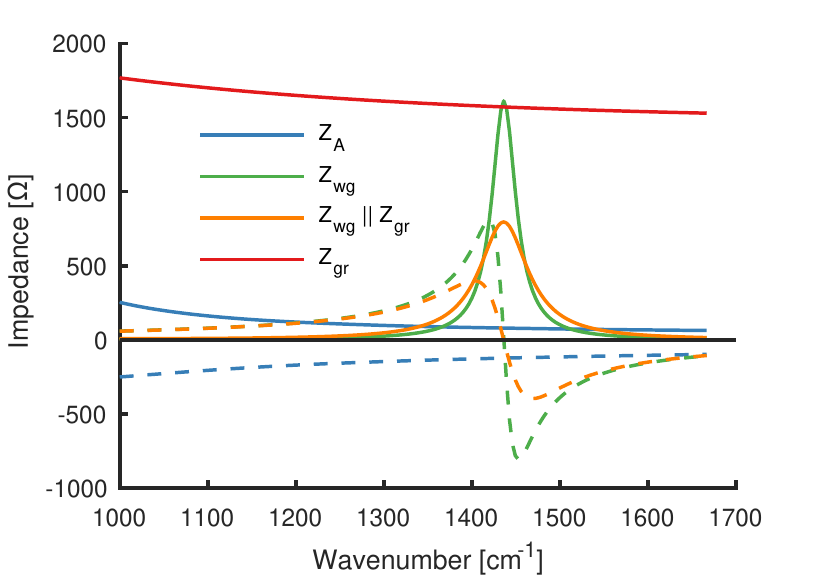} }}
\qquad
\subfloat{{\includegraphics[width=0.65\textwidth]{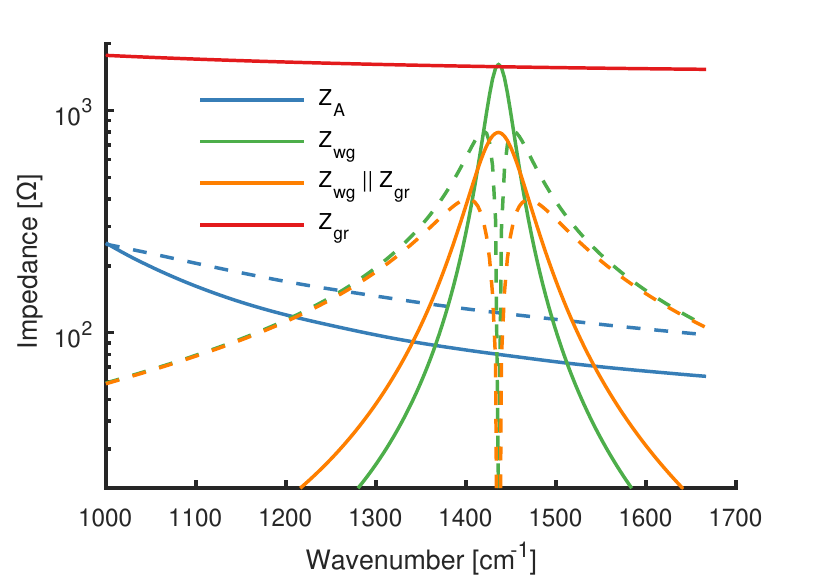} }}
\caption{Wavenumber dependence of the impedances in the circuit in Main Figure 2a.  Above: Linear scaling. Below: Log-abs scaling. The antenna featured here has dimensions $d = 2.0\,\um$, $h = 5.5\,\um$, and $w = 0.4\,\um$. }
\label{impedances}
\end{figure}

\newpage
\subsubsection*{Graphene absorption spectra and individual antenna contributions for metasurfaces of varied pitch}

\begin{figure}[h!]\centering
\includegraphics[width=0.92\textwidth]{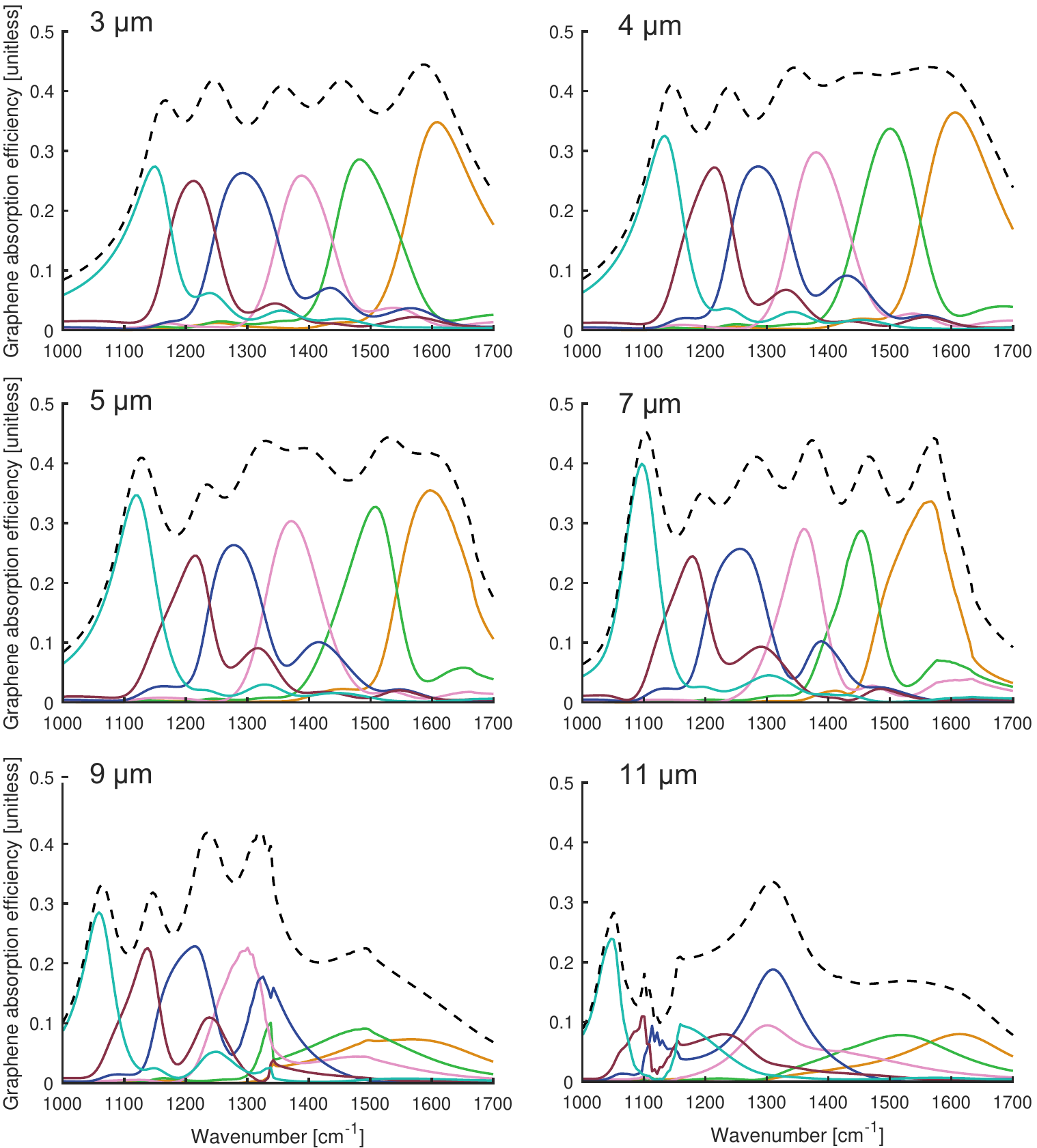}
\caption{Graphene absorption efficiency and individual antenna contributions for metasurfaces of varied pitch. Full unit cell dimensions for each case are given in Main Figure 6.}
\label{pitchvar}
\end{figure}

\newpage
\subsubsection*{Antenna radiation pattern and off-angle excitation specular reflection condition}

\begin{figure}[h!]\centering
\includegraphics[width=\textwidth]{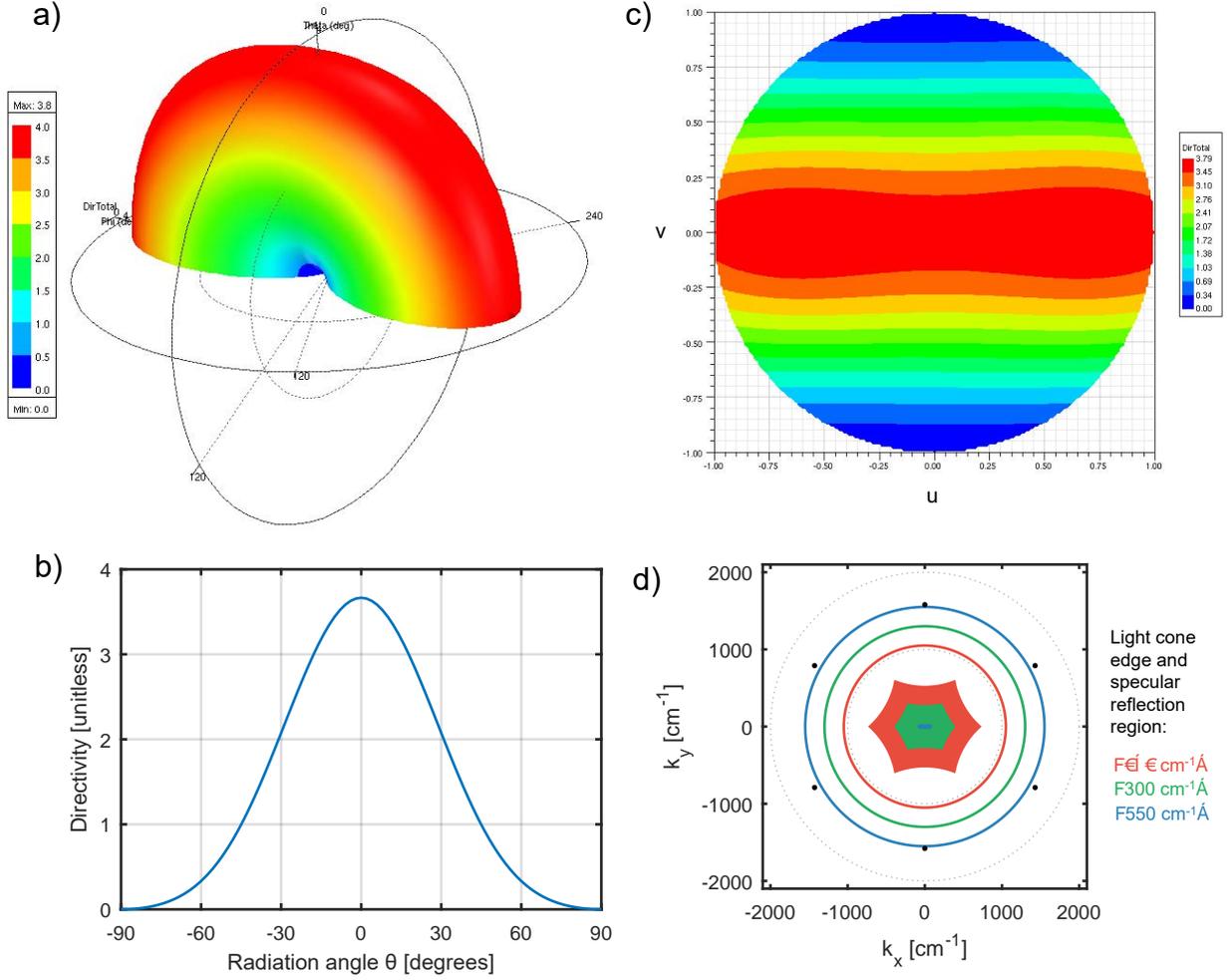}
\caption{Different representations of the on-resonance far-field radiation pattern of a $6.5\,\um$-long slot antenna, which is qualitatively very similar to those of antennas of other lengths. a) 3D-representation of the far-field directivity pattern. $D\left(\theta,\phi\right)$ goes to zero for light incident along the long axis of the aperture. b) Directivity vs. elevation angle $\theta$ along the azimuth of the antenna's long axis, $\phi=90^\circ$. c) Directivity in directional cosine space. Note also that $k_x=k_0\,u$ and $k_y=k_0\,v$ where $k_x$ and $k_y$ are the $x$- and $y$-components of the incident wavevector. d) Depiction in $k_x\,k_y$--space of the light cone edges (colored rings) and the specular reflection-only incident light directions (colored patches) for three different wavenumbers for the $7\,\um\times 12.667\,\um$ 6-antenna unit cell design in Main Figure 1a. The black dots represent the reciprocal lattice points of the metasurface. For each wavelength, the specular reflection-only region is the set of remaining $\{k_x,k_y\}$ after subtracting from the light cone all copies of the light cone transposed by all nonzero reciprocal lattice vectors.}
\label{radiation}
\end{figure}

\newpage
\subsubsection*{Off-angle absorption spectra for simplified, perfectly conducting metasurface absorber}

\begin{figure}[h!]\centering
\includegraphics[width=\textwidth]{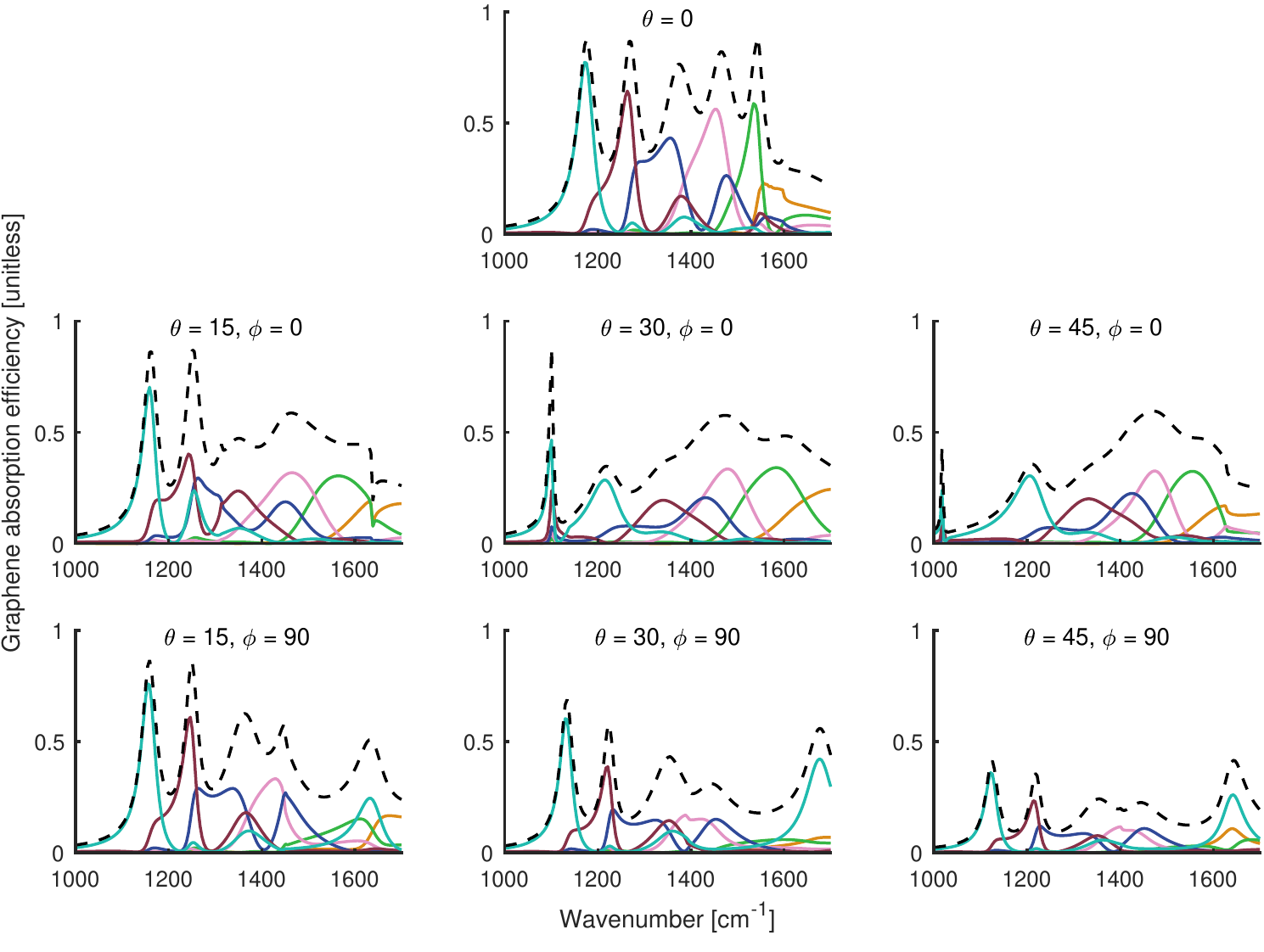}
\caption{Graphene absorption efficiency and individual antenna contributions for a perfectly conducting metasurface with various off-angle excitation directions. Using Perfect Electrical Conductor (PEC) as the metasurface material allows us to use a coarser mesh of $40\,\nm$, as the electric field does not penetrate the conductor (the ``skin depth'' which requires careful modelling for realistic metal). We also adjust the metasurface dimenisons to align all features to the mesh: The unit cell is $7.2\times 12.64\,\um$, and the antenna lengths are 3.4, 3.8, 4.4, 5.04, 5.8, and $7.36\,\um$. The topmost plot shows the absorption spectra for normally incident light. The second row depicts the case of light incident perpendicular to the antennas' long axes (along the line of maximum directivity), and the third row depicts light incident perpendicular to the antennas' short axes.}
\label{offangle}
\end{figure}

\newpage
\subsubsection*{Graphene absorption spectra and individual antenna contributions for metasurfaces with different numbers of antennas}

\begin{figure}[h!]\centering
\includegraphics[width=0.92\textwidth]{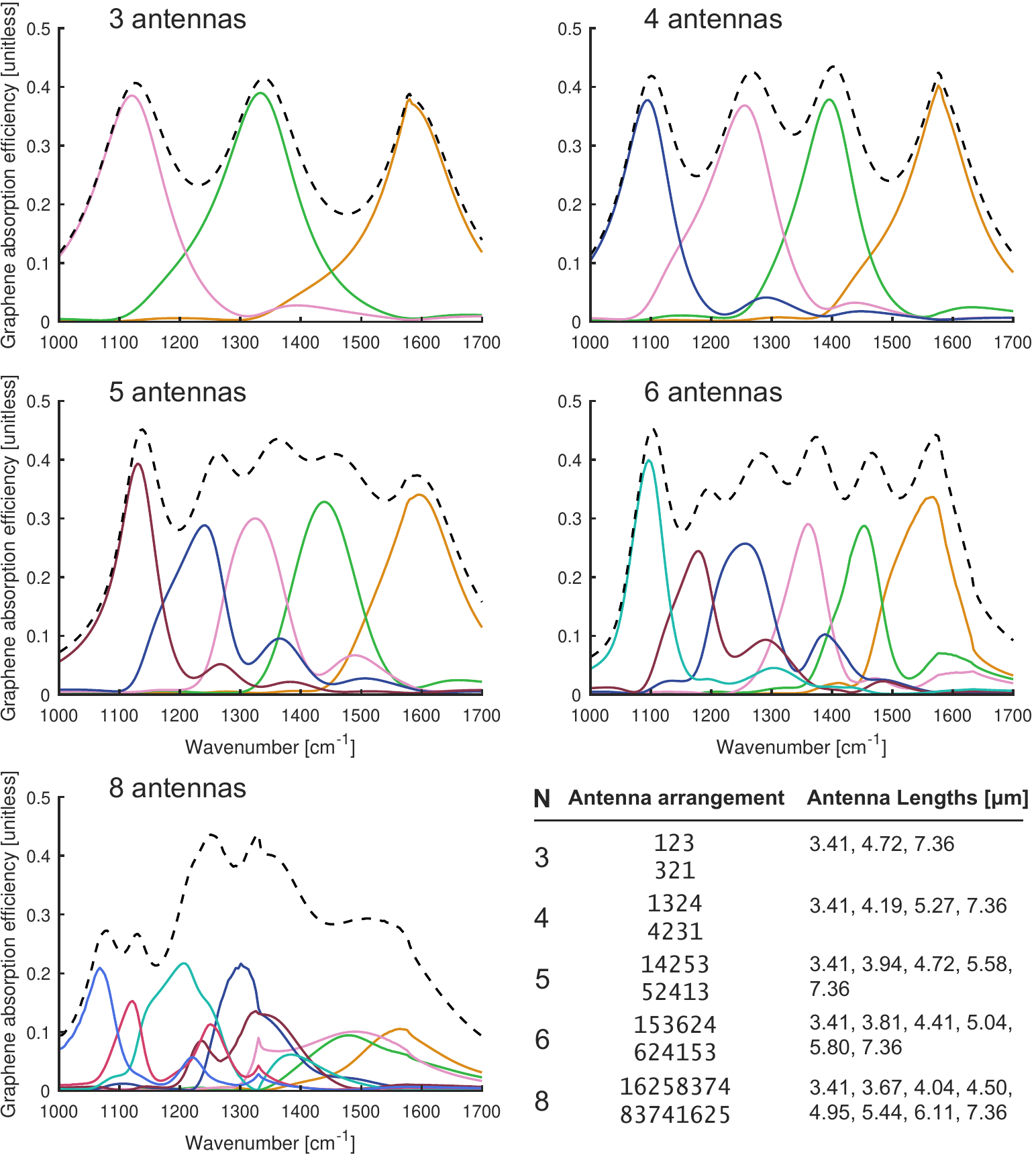}
\caption{Graphene absorption efficiency and individual antenna contributions for metasurfaces with varied numbers $N$ of antennas of different lengths. The lower-right table shows the arrangements of the antennas within the unit cell, with $1$, $2$...$N$ corresponding to the shortest, second shortest, etc. up to longest antenna. The table also lists the lengths of the antennas in each case, chosen roughly to maintain a constant degree of frequency-space overlap between adjacent spectral channels as simulated on an individual antenna basis. Antennas are placed on a $1.167\,\um\times 6.333\,\um$ grid as in Main Figure 5.}
\label{pitchvar}
\end{figure}

\newpage
\subsubsection*{Graphene absorption spectra and individual antenna contributions for metasurfaces with different FDTD mesh}

\begin{figure}[h!]\centering
\includegraphics[width=0.7\textwidth]{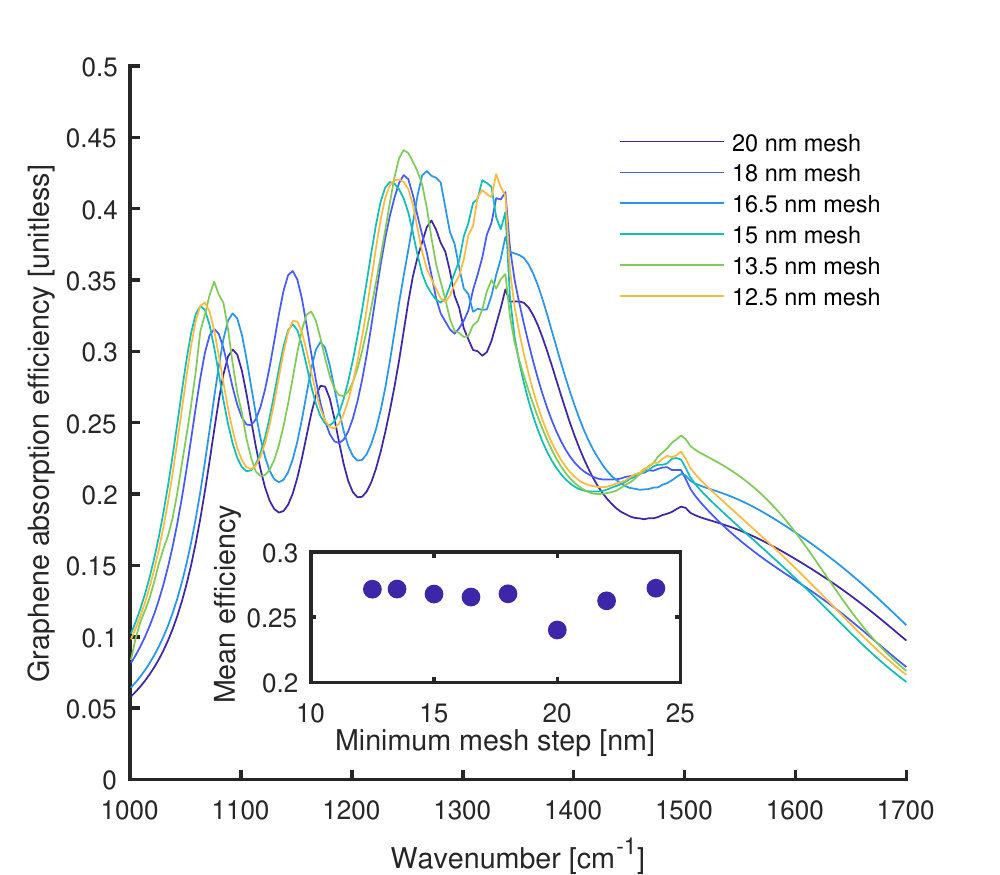}
\caption{Graphene absorption efficiency and individual antenna contributions for a $9\,\um$ by $13.333\,\um$ metasurface with varied minimum FDTD mesh size as discussed in the Methods section of the main paper. The inset plots the mesh dependence of the mean graphene absorption efficiency in the $1050\,\cm^{-1}$ to $1600\,\cm^{-1}$ range.}
\label{meshvarr}
\end{figure}

\newpage
\subsubsection*{Graphene absorption spectra and individual antenna contributions for a metasurface with graphene doped to 0.05 eV}

\begin{figure}[h!]\centering
\includegraphics[width=0.65\textwidth]{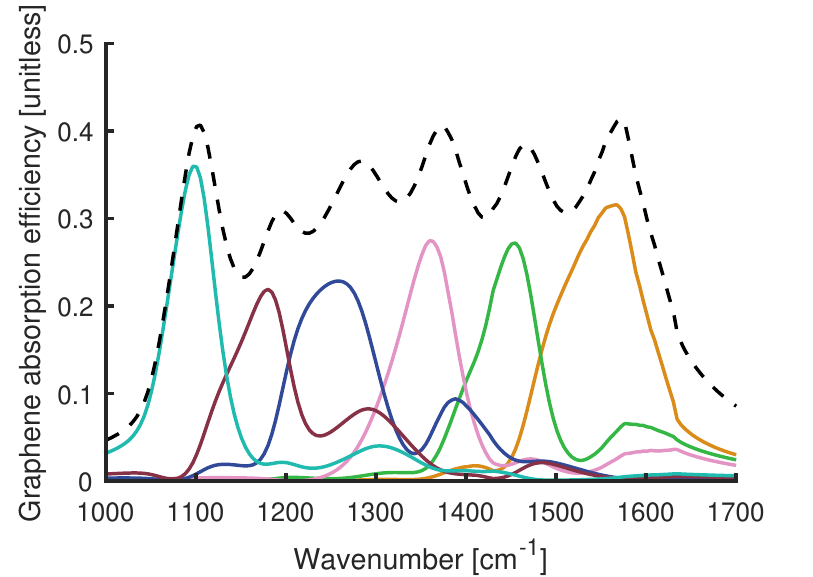}
\caption{Graphene absorption efficiency and individual antenna contributions for a $7\,\um$ by $12.667\,\um$ metasurface with the same geometric parameters used to generate Main Figure 5, but with the graphene doped to 0.05 eV which decreases the graphene sheet conductivity to a degree, especially for longer wavelengths. Here, the mean efficiency averaged between $1050\,\cm^{-1}$ and $1600\,\cm^{-1}$ is $33\%$. This doping level is chosen to maximize the Seebeck coefficient for the graphene model used to approximate the device detectivity as discussed in the main text.}
\label{antennas6_0eV05}
\end{figure}